# Helmholtz wave trajectories
# in classical and quantum physics

A. Orefice*, R. Giovanelli and D. Ditto

*University of Milan*
*Di.Pro.Ve. - Via G. Celoria, 2 - 20133 - Milano (Italy)*

**Abstract** - The behavior of classical and quantum wave beams in stationary media is shown to be ruled by a "Wave Potential" function encoded in Helmholtz-like equations, determined by the structure itself of the beam and taking, in the quantum case, the form of Bohm's "Quantum Potential", which is therefore not so much a "quantum" as a "wave" property. Exact, deterministic motion laws, mutually coupled by this term and describing wave-like features such as diffraction and interference, are obtained in terms of well defined Hamiltonian trajectories, and shown to reduce to the laws of usual geometrical optics and/or classical dynamics when this coupling term is neglected.

As far as the *quantum* case is concerned, the approach proposed in the present paper, suggested by the direct extension of the treatment holding for *classical* waves, describes the motion of classical-looking, point-like particles, without resorting to the use of travelling wave-packets.

## 1 Introduction

The present paper is concerned with the description in terms of trajectories of typically wave-like features such as diffraction and interference, and aims at developing a simple and exact common approach for *classical* and *quantum* wave beams. As we shall see, indeed, the trajectories pertaining to classical and quantum waves may be submitted to quite similar treatments.

While a trajectory-based approach going beyond the geometrical optics approximation is quite unusual (with few exceptions [1,2]) in the case of *classical waves*, it is increasingly frequent for *quantum matter waves* - after the original formulation of the de Broglie-Bohm theory (dBBt) in terms of a generalized Hamilton-Jacobi equation including a suitably defined *Quantum Potential* [3-6] - ever since the pioneering approaches of **Refs.[7-9]**. Ambitious systematizations of the dBBt were presented, in the meanwhile, in **Refs.[10,11]**.

Two main approaches may be distinguished, in the last decade, in this flourishing field.

The first one [12-17] is based on hydrodynamic-like equations, where the trajectories mark the flow-lines of quantum probability, and the Quantum Potential is obtained from a solution of the time-dependent Schrödinger equation, mainly based on the Gaussian wave-packets of Heller's approach [18]. Applications to problems involving, for instance, atom scattering by metal surfaces and cold neutron diffraction by an arbitrary number of slits are treated by means of appropriate numerical techniques.

---

* e-mail: adriano.orefice@unimi.it

The second line of research **[19-21]** builds the Quantum Potential by means of an iterative solution of the time dependent quantum Hamilton-Jacobi equation, making use of an infinite set of equations which describe fully local complex quantum trajectories. This procedure is applied, for instance, to the numerical description of the one-dimensional scattering and tunneling of Gaussian wave-packets. A complex quantum Hamilton-Jacobi formalism is also employed in **Refs.[22,23]** for the analysis of wave-packet interference.

It is finally worth mentioning two important collections of papers, **Refs.[24,25]** (the second of which includes a contribution by the present Authors), where applications of quantum trajectories to many different cases, from nano-technologies to cosmology, are performed and analyzed.

Our approach to a trajectory-based description of wave-like features in stationary media makes use of a simple and exact formalism which is shown to hold *both for classical and quantum waves*, and does not require, in the quantum case, the solution, step by step, of a time-dependent Schrödinger or Hamilton-Jacobi equation.

In Sect.2 an exact set of Hamiltonian equations is deduced from the Helmholtz equation - to our knowledge, *for the first time* - for *classical* waves, allowing the treatment of general wave-like features such as diffraction and interference. A (stationary) *Wave Potential* function, due to the structure itself of the Helmholtz equation, is defined and shown to couple all the rays of a wave beam in a kind of self-refractive effect.

In Sect.3, thanks to the (well known) observation that the time-independent Schrödinger equation is itself a Helmholtz-like equation, the Hamiltonian procedure of Sect. 2 is extended to *quantum* matter waves, showing that the so-called Quantum Potential is *a particular case of Wave Potential*, and is therefore *not so much a "quantum" as a "wave" property*. Each particle is not only conceived in principle, but described in practice, as a point-like object, endowed with a well defined position and momentum and not requiring the probabilistic or statistical use of travelling wave-packets.

Sect.4 contains examples of numerical integration of our Hamiltonian ray tracing system, in a form holding both for classical and quantum waves, in a number of diffraction and/or interference cases.

Sect.5 contains a discussion of our basic assumptions.

Sect.6 summarizes the goals reached by means of this approach.

## 2    The case of classical waves

We assume here both *stationary media* (allowing the best theoretical and experimental analysis of diffraction and/or interference patterns) and *wave mono-chromaticity*, warranting the very possibility of defining the concept of "trajectories", i.e. of lines orthogonal to monochromatic wave-fronts. Although our considerations may be easily developed for many kinds of *classical* (from acoustic to seismic) waves, we shall refer in the present Section, in order to fix ideas, to a classical *electromagnetic* wave beam travelling through a stationary, isotropic and (generally) inhomogeneous  dielectric medium  according  to a scalar wave  equation of the simple form **[26]**

$$\nabla^2 \psi - \frac{n^2}{c^2} \frac{\partial^2 \psi}{\partial t^2} = 0 \quad , \tag{1}$$

where $\psi(x, y, z, t)$ represents any component of the electric and/or magnetic field, $n(x, y, z)$ is the (time independent) refractive index of the medium and $\nabla^2 \equiv \frac{\partial^2}{\partial x^2} + \frac{\partial^2}{\partial y^2} + \frac{\partial^2}{\partial z^2}$ . By assuming

$$\psi(x, y, z, t) = u(x, y, z) e^{-i\omega t} \quad , \tag{2}$$

with obvious definition of $u(x, y, z)$ and $\omega$ , we get the Helmholtz equation

$$\nabla^2 u + (n k_0)^2 u = 0 \quad , \tag{3}$$

where $k_0 \equiv \frac{2\pi}{\lambda_0} = \frac{\omega}{c}$ . Notice that, limiting here our considerations to the case of monochromatic waves, we did not explicitly mention (for simplicity sake) the possible dependence of $\psi$, $u$ and $n$ on $\omega$ .

If we now perform the quite general and well-known replacement

$$u(x, y, z) = R(x, y, z) e^{i \varphi (x, y, z)} \quad , \tag{4}$$

with real $R(x,y,z)$ and $\varphi (x,y,z)$, and separate the real from the imaginary part, eq.(3) splits into the coupled system **[26]**

$$\begin{cases} (\vec{\nabla} \varphi)^2 - (n k_0)^2 = \dfrac{\nabla^2 R}{R} \\ \vec{\nabla} \cdot (R^2 \vec{\nabla} \varphi) = 0 \end{cases} \tag{5}$$

where $\vec{\nabla} \equiv \partial / \partial \vec{r} \equiv (\partial / \partial x, \partial / \partial y, \partial / \partial z)$ and $\vec{r} \equiv (x, y, z)$ .

The *second* of eqs. (5) expresses the constancy of the flux of the vector $R^2 \vec{\nabla} \varphi$ along any tube formed by the field lines of the *wave vector*

$$\vec{k} = \vec{\nabla} \varphi . \tag{6}$$

As far as the *first* of eqs.(5) is concerned, we multiply it, for convenience, by the constant factor $\dfrac{c}{2 k_0}$ , thus obtaining, by means of eq.(6), the relation

$$D(\vec{r}, \vec{k}) \equiv \frac{c}{2 k_0} [k^2 - (n k_0)^2 - \frac{\nabla^2 R}{R}] = 0 \quad , \tag{7}$$

whose differentiation

$$\frac{\partial D}{\partial \vec{r}} \cdot d\vec{r} + \frac{\partial D}{\partial \vec{k}} \cdot d\vec{k} = 0 \quad , \tag{8}$$

with $\partial / \partial \vec{k} \equiv (\partial / \partial k_x, \ \partial / \partial k_y, \ \partial / \partial k_z)$, immediately provides a Hamiltonian ray-tracing system of the form

$$\begin{cases} \dfrac{d\vec{r}}{dt} = \dfrac{\partial D}{\partial \vec{k}} = \dfrac{c\,\vec{k}}{k_0} \\[2mm] \dfrac{d\vec{k}}{dt} = -\dfrac{\partial D}{\partial \vec{r}} = \vec{\nabla} \left[ \dfrac{c\,k_0}{2} n^2(x,y,z) - W(x,y,z) \right] \end{cases} \tag{9}$$

where

$$W(x,y,z) = -\frac{c}{2k_0} \frac{\nabla^2 R}{R} \tag{10}$$

and a ray velocity $\vec{v}_{ray} = \dfrac{c\,\vec{k}}{k_0}$ is implicitly defined. It is easily seen that, as long as $k \equiv \left| \vec{k} \right| = k_0$, we'll have $v_{ray} \equiv \left| \vec{v}_{ray} \right| = c$ . The function $W(x,y,z)$, which we define in eq. (10) and call "Helmholtz Wave Potential", couples the geometry and motion laws of the rays of the beam in a kind of self refraction, strongly affecting their propagation. Such a term (which has the dimensions of a *frequency*) represents an *intrinsic property encoded in the Helmholtz equation itself*, and is determined by the structure of the beam. We observe, from the second of eqs.(5), that

$$\vec{\nabla}_\bullet (R^2 \, \vec{\nabla} \, \varphi) \equiv 2\, R\, \vec{\nabla} R \cdot \vec{\nabla}\varphi + R^2\, \vec{\nabla} \cdot \vec{\nabla}\varphi = 0. \tag{5'}$$

Since no new trajectory may suddenly arise in the space region spanned by the beam, we must have $\vec{\nabla} \cdot \vec{\nabla}\varphi = 0$, so that $\vec{\nabla} R \cdot \vec{\nabla}\varphi = 0$: the amplitude R (as well as its functions and derivatives) is distributed, at any time, on the wavefront reached at that time, so that both $\vec{\nabla} R$ and $\vec{\nabla} W$ are perpendicular to $\vec{k} \equiv \vec{\nabla}\varphi$ . A basic, *particular* consequence of this *general* property is the fact that the *absolute value* of the ray velocity remains (in the particular case of electromagnetic wave propagation *in vacuo*) equal to *c* all along each ray trajectory, because such a perpendicular term may only modify the *direction*, but not the *amplitude*, of the wave vector $\vec{k}$ . The only possible changes of $k \equiv \left| \vec{k} \right|$ may be due, in a medium different from vacuum, to its refractive index $n(x,y,z)$.

The knowledge of the distribution of R on a wave-front is the necessary and sufficient condition to determine its distribution on the next wave-front. Thanks to the constancy of the flux of $R^2\, \vec{\nabla}\varphi$ , in fact, the function $R(x,y,z)$, once assigned on the surface from which the beam is assumed to start, may be built up step by step, together with the Wave Potential $W(x,y,z)$, along the ray trajectories.

This allows the numerical integration **[27,28]** of the Hamiltonian system (9), and provides **both** an exact *stationary* "weft" of coupled "rails" (which we could call "Helmholtz trajectories"), along which the rays are channeled, **and** the ray motion laws along them, starting (with an assigned wave-vector) from a definite point of the launching surface and coupled by the Wave Potential $W(x,y,z)$.

Let us observe that when, in particular, the space variation length $L$ of the beam amplitude $R(x,y,z)$ satisfies the condition $k_0 L \gg 1$, the **first** of eqs.(5) is well approximated by the *eikonal equation* **[26]**

$$(\vec{\nabla} \varphi)^2 \simeq (n k_0)^2 \quad , \tag{11}$$

decoupled from the **second** equation, and the term containing the wave potential $W(x,y,z)$ may be dropped from the ray tracing system (9). In this *eikonal* (or "*geometrical optics*") *approximation* the rays are not coupled by the Wave Potential, and propagate independently from one another.

Let us recall here, moreover, that a quasi-optical, trajectory-based Hamiltonian treatment of the injection and propagation of electromagnetic Gaussian beams at the electron-cyclotron resonant frequency in the toroidal thermonuclear plasmas of Tokamaks such as JET and FTU was presented in 1993/94 by one of the Authors (A.O., **[29,30]**). A complex eikonal equation, amounting to a first order approximation of the beam diffraction, was adopted in order to overcome the collapse of the ordinary geometrical optics approximation.

Coming back to the most general case, let us finally observe that if we pass to *dimensionless variables* by expressing

- the space variable $\vec{r}$ (together with the space operators $\vec{\nabla}$ and $\nabla^2$) in terms of a physical length $w_0$ (to be defined later on),

- the wave vector $\vec{k}$ in terms of $k_0$, and

- the time variable $t$ in terms of $w_0 / c$,

and maintaining for simplicity the names $\vec{r}, \vec{k}, t$, the Hamiltonian system (9) takes on the dimensionless form

$$\begin{cases} \dfrac{d\vec{r}}{dt} = \vec{k} \\[2mm] \dfrac{d\vec{k}}{dt} = \dfrac{1}{2} \vec{\nabla} \, [\, n^2 + \left(\dfrac{\varepsilon}{2\pi}\right)^2 \, G(x,y,z)] \end{cases} \tag{12}$$

where

$$\varepsilon = \lambda_0 / w_0 \tag{13}$$

and the Wave Potential (with opposite sign) is represented by the (dimensionless) function

$$G(x,y,z) = \frac{\nabla^2 R}{R} \quad . \tag{14}$$

Notice that different values of $\varepsilon \equiv \lambda_0 / w_0$ (i.e. *different frequencies* $\omega = 2\pi c / \lambda_0$, for a fixed value of the assumed unit of length, $w_0$) lead to different values of the coefficient weighting the effect of the potential function G, and therefore to *different trajectories*. In this sense we may speak of a *dispersive* character of the *Wave Potential* itself.

## 3 The case of quantum (matter) waves

Let us pass now to the case of a mono-energetic beam of non-interacting particles of mass $m$ launched with an initial momentum $\vec{p}_0$ into a force field deriving from a potential energy $V(x,y,z)$ not explicitly depending on time. The *classical* motion of each particle of the beam may be described, as is well known, by the time-independent Hamilton-Jacobi equation [26]

$$(\vec{\nabla} S)^2 = 2\,m\,[E - V(x,y,z)] \quad , \tag{15}$$

where $E = p_0^2 / 2m$ is the total energy of the particle, and the basic property of the function $S(x,y,z)$ is that the particle momentum is given by

$$\vec{p} = \vec{\nabla} S \quad . \tag{16}$$

The analogy between eqs.(15), (16) on the one hand, and eqs.(11), (6) on the other, together with an illuminating comparison between Fermat's and Maupertuis' variational principles, suggested to de Broglie [31] and Schrödinger [32,33] , as is well known [34], that the *classical* particle dynamics could be the *geometrical optics approximation* of a more general *wave-like reality* described by a suitable *Helmholtz-like* equation. Such an equation is immediately obtained, indeed, from eq.(3), by means of the replacements

$$\begin{cases} \varphi = \dfrac{S}{a} \\[2mm] \vec{k} \equiv \vec{\nabla}\,\varphi \;=\; \dfrac{\vec{\nabla} S}{a} \;=\; \dfrac{\vec{p}}{a} \\[2mm] k_0 \equiv \dfrac{2\pi}{\lambda_0} = \dfrac{p_0}{a} \equiv \dfrac{\sqrt{2mE}}{a} \\[2mm] n^2(x,y,z) = 1 - \dfrac{V(x,y,z)}{E} \end{cases} \tag{17}$$

directly inspired by the afore-mentioned analogy. The parameter *"a"* represents a constant *action* whose value is *a priori* arbitrary, but whose choice

$$a = \hbar \cong 1.0546 \times 10^{-27} \, erg \times s \tag{18}$$

is suggested by the de Broglie's Ansatz **[31]**

$$\vec{p} = \hbar \, \vec{k} \, , \tag{19}$$

thus transforming eq.(3) into the standard *time-independent* Schrödinger equation holding in a stationary field $V(x,y,z)$

$$\nabla^2 u + \frac{2m}{\hbar^2} [E - V(x,y,z)] \, u = 0 \quad . \tag{20}$$

From eqs.(2) and (20) we get, on the one hand, the relation

$$\nabla^2 \psi - \frac{2m}{\hbar^2} \, V \, \psi = -\frac{2m}{\hbar^2} \, E \, \psi \equiv -\frac{2m \, i}{\hbar} \, \frac{E}{\hbar \, \omega} \, \frac{\partial \psi}{\partial t} \tag{21}$$

which, by assuming (with de Broglie and Schrödinger **[31-33]** ) the Planck relation

$$E = \hbar \omega, \tag{22}$$

i.e. by attributing to the energy of a material particle a relation coming, *stricto sensu*, from the radiation theory - takes on the standard form of the (*intrinsically complex*) *time-dependent* Schrödinger equation in a stationary field $V(x,y,z)$,

$$\nabla^2 \psi - \frac{2m}{\hbar^2} \, V(x,y,z) \, \psi = -\frac{2m \, i}{\hbar} \, \frac{\partial \psi}{\partial t} \quad , \tag{23}$$

where $E$ and $\omega$ are not involved. Let us observe, on the other hand, that, just like the Helmholtz equation (3) is associated with the wave equation (1), the *Helmholtz-like* eq.(20) is associated - *via* eq.(2) - with the *ordinary-looking* wave equation **[34]**

$$\begin{aligned} \nabla^2 \psi &= \frac{2m}{(\hbar\omega)^2} (E - V) \, \frac{\partial^2 \psi}{\partial t^2} \equiv \frac{2m}{E^2} (E - V) \, \frac{\partial^2 \psi}{\partial t^2} \\ &\equiv \frac{2m}{(\hbar\omega)^2} (\hbar\omega - V) \, \frac{\partial^2 \psi}{\partial t^2} \end{aligned} \tag{24}$$

providing significant information about the wave propagation and its dependence on frequency and/or energy, and therefore its dispersive character.

By applying now to the *Helmholtz-like* eq.(20) the same procedure leading from eq.(3) to eqs.(5), and assuming therefore

$$u(x,y,z) = R(x,y,z) \, e^{\,i\,S(x,y,z)/\hbar} \quad , \quad (25)$$

eq.(20) splits into the coupled system **[35]**

$$\begin{cases} (\vec{\nabla} S)^2 - 2m(E - V) = \hbar^2 \dfrac{\nabla^2 R}{R} \\ \vec{\nabla} \cdot (R^2 \, \vec{\nabla} S) = 0 \end{cases} \quad , \quad (26)$$

analogous to eqs (5). By simply maintaining eq.(16), the first of eqs. (26) may be written in the form of a generalized, time-independent Hamiltonian

$$H \, (\, \vec{r}, \, \vec{p} \,) \equiv \frac{p^2}{2m} + \; V(x,y,z) + Q(x,y,z) = \; E \qquad (27)$$

where the function

$$Q(x,y,z) = -\frac{\hbar^2}{2m} \frac{\nabla^2 R}{R} \;\;, \qquad (28)$$

(which has the dimensions of an *energy*) is structurally analogous to the Wave Potential function $W(x,y,z)$ of eq.(10), and turns out to coïncide with the well known *Quantum Potential* of the de Broglie-Bohm theory **[3-11].** Such a term is clearly due not so much to the "quantum" behavior of the particles as to their "wave-like" nature, suggested by de Broglie and Schrödinger. By differentiating eq. (27) we get the relation

$$\frac{\partial \, H}{\partial \, \vec{r}} \cdot d \, \vec{r} + \frac{\partial \, H}{\partial \, \vec{p}} \cdot d \, \vec{p} = 0 \qquad (29)$$

with $\partial / \partial \vec{p} \equiv (\partial / \partial p_x, \; \partial / \partial p_y, \; \partial / \partial p_z)$, leading to a Hamiltonian dynamical system of the form

$$\begin{cases} \dfrac{d \, \vec{r}}{d \, t} = \dfrac{\partial \, H}{\partial \, \vec{p}} = \dfrac{\vec{p}}{m} \\ \dfrac{d \, \vec{p}}{d \, t} = -\dfrac{\partial \, H}{\partial \, \vec{r}} = -\vec{\nabla}[V(x,y,z) + Q(x,y,z)] \end{cases} \qquad (30)$$

This *quantum* dynamical system is strictly similar to the exact, deterministic ray-tracing system (9) concerning *classical* electromagnetism. In spite of its "quantum" context, therefore, we shall submit it to the same interpretation and mathematical treatment applied in the previous (classical) case. The presence of the potential $Q(x,y,z)$ causes, once more, the "*Helmholtz coupling*" of the geometry and motion laws of the rays of the whole beam, and its absence or omission would reduce the *quantum* system (30) to the standard *classical* set of dynamical equations, which constitute therefore, as expected, its *geometrical optics approximation*. In complete analogy with the classical electromagnetic case of the previous Section,

1) the term $-\vec{V} Q(x,y,z)$ (behaving here as a force) is perpendicular to $\vec{p} \equiv \vec{V} S$, so that it cannot modify the *amplitude* of the particle momentum (while modifying, in general, its *direction*), and the only possible amplitude changes of $\vec{p}$ could be due to the presence of an external potential $V(x,y,z)$: in other words, *no energy exchange may ever occur between particles and quantum potential*;

2) the relations $\vec{p} = \vec{V} S$ and $\vec{V} \cdot (R^2 \vec{V} S) = 0$ allow to obtain step by step, along the particle trajectories, both $R(x,y,z)$ and $Q(x,y,z)$, thus avoiding the solution of the time-dependent Schrödinger equation involved by Bohm's suggestion to make use of travelling wave-packets [4], i.e. of statistical ensembles representing a practical necessity, but not a manifestation of an inherent lack of determination of the particle nature and motion.

We stick, in other words, *to the spirit, and not to the letter*, of Bohm's interpretation of the Quantum Theory in terms of precisely definable and continuously varying values of "hidden" variables - such as position and momentum - determining the complete behavior of individual particles. The dynamic Hamiltonian system (30) provides the exact, complete, deterministic description of *classical-looking, point-like particles* starting from assigned point-like positions on the initial wave-front (with an assigned launching momentum) and following well defined *stationary* trajectories, without importing the uncertainty involved by a *wave-packet* representation. Particles of the beam starting from different points of the launching surface move along *stationary* trajectories coupled *ab initio* by the Quantum Potential.

In complete analogy, moreover, with the previous electromagnetic case, the quantum Hamiltonian system (30) may be put in a suggestive *dimensionless* form by expressing lengths (as well as $\vec{V}$ and $\nabla^2$) in terms of a physical length $w_0$ ( to be defined later on), momentum in terms of $p_0$ and time in terms of $w_0 / v_0$, with $v_0 = p_0 / m$:

$$\begin{cases} \dfrac{d\,\vec{r}}{d\,t} = \vec{p} \\ \dfrac{d\,\vec{p}}{d\,t} = \dfrac{1}{2} \vec{V} \,[\, -\dfrac{V}{E} + (\dfrac{\varepsilon}{2\pi})^2 \; G(x,y,z)]\end{cases} \tag{31}$$

where the parameter $\varepsilon$ and the (dimensionless) potential function $G(x,y,z)$ are given, once more, by eqs. (13) and (14) [1]. Not surprisingly, the *quantum* system (31) turns out to coïncide with the *classical* dimensionless system (12) by simply replacing $\vec{k}$ by $\vec{p}$ and $n^2$ by *(1-V/E)*, in agreement with eqs.(17). The coupling due to $G(x,y,z)$ is therefore a physical phenomenon affecting *both classical and quantum waves*, and its absence would reduce the relevant equations to the ones, respectively, of standard geometrical optics and of classical dynamics.

---

[1] A unidimensional form of the function $G$ is recognized to be analogous to the *quantum potential function*, and called *acoustical potential*, in **Refs. [36,37]**, where the similarity between Webster's horn equation and the Klein-Gordon equation is stressed and applied to the reconstruction of the geometry of an acoustical duct from the radiated wave.

Let us observe once more that different values of $\varepsilon \equiv \lambda_0 / w_0$ (i.e. different frequencies $\omega = 2\pi c / \lambda_0$, for a fixed value of the assumed unit of length, $w_0$) lead to different sets of trajectories.

## 4 Numerical examples

Once assigned on the launching surface of the beam, the wave amplitude profile $R(x,y,z)$ and the consequent potential function $G(x,y,z)$ may be numerically built up step by step, together with their derivatives, along the beam trajectories, making use of eqs. (5) and/or (26). We present here some applications of the Hamiltonian systems (12) and/or (31) to the propagation of collimated beams injected at $z=0$, parallel to the $z$-axis, simulating wave *diffraction* and/or *interference* through suitable slits, each one of half width $w_0$. Here we perform, therefore, the choice of the physical meaning of this length, and we assume $\varepsilon \equiv \lambda_0 / w_0 \ll 1$.

The problem is faced by taking into account, for simplicity sake, either (*quantum*) particle beams in the absence of external fields (*V = 0*) or (*classical*) electromagnetic beams *in vacuo* $(n^2 = 1)$, with a geometry allowing to limit the computation to the $(x,z)$-plane. Because of the coïncidence between the (dimensionless) Hamiltonian systems (12) and (31), the only choice to be performed is between the variable names $\vec{k}$ or $\vec{p}$ - and we opt here for the second one, reminding however that we are not necessarily speaking of quantum topics. Recalling that, because of the transverse nature of the gradient $\vec{V}G$, the *amplitude* of $\vec{p}$ remains unchanged (in the absence of external fields and/or refractive effects) along each trajectory, we have

$$p_x(t=0) = 0; \ \ p_z(t=0) = 1$$
$$p_z(t \geq 0) = \sqrt{1 - p_x^2(t \geq 0)}$$

(32)

and the dimensionless Hamiltonian system (31) reduces to the form

$$\begin{cases} \dfrac{d\,x}{d\,t} = \ p_x \\[2mm] \dfrac{d\,z}{d\,t} = \ \sqrt{1 - p_x^2} \\[2mm] \dfrac{d\,p_x}{d\,t} = \ \dfrac{\varepsilon^2}{8\,\pi^2} \ \ \dfrac{\partial\,G(x,z)}{\partial\,x} \end{cases}$$

(33)

where

$$G(x,z) \equiv \frac{\nabla^2 R}{R} = \frac{\partial^2 R / \partial x^2}{p_z^2 \, R} \quad . \qquad (34)$$

We assumed throughout the present computations the value $\varepsilon \equiv \lambda_0 / w_0 = 1.65 \times 10^{-4}$. Let us mention, for comparison, that a case of cold neutron diffraction was considered in **Ref.[13]** with

$$\lambda_0 = 19.26 \times 10^{-4} \mu m \, , \;\; 2w_0 = 23 \mu m \, ,$$

$$\varepsilon = \lambda_0 / w_0 \cong 1.67 \times 10^{-4}$$

The analysis of the fringeless diffraction of a beam whose launching amplitude profile is a single Gaussian

$$R(x; z = 0) = exp(-x^2) \qquad (35)$$

was thoroughly treated in **Ref.[27]**, and shall not be repeated here. We limit ourselves to compare in Fig.1 (in order to show the *dispersive* effect of the potential *G*) the *waist* trajectories **[38]**

$$x = \pm \sqrt{1 + (\frac{\varepsilon \, z}{\pi})^2} \qquad (36)$$

of the Gaussian beams obtained for two different values of $\varepsilon$ :

$\varepsilon_1 = 1.65 \times 10^{-4}$ (continuous lines) and $\varepsilon_2 = 1.25 \times 1.65 \times 10^{-4}$ (dashed lines),

*and exactly coïnciding with the corresponding numerical trajectories* starting from $x = \pm 1$. It's worth while reminding that lengths are measured in terms of $w_0$. The beam launching amplitude distribution $R(x; z = 0)$ (from whose normalization the function *G* is obviously independent) is assigned, in the following, by means of two different models consisting of suitable superpositions of Gaussian functions either in the form

$$R(x; z = 0) = a \, exp(-q^2 x^2) + b \sum_{N=1}^{M} \left\{ exp\left[ -q^2 \left( x - N x_C \right)^2 \right] + exp\left[ -q^2 \left( x + N x_C \right)^2 \right] \right\} \quad (37a)$$

or in the form

$$R(x; z = 0) = \sum_{N=-M}^{M} \left\{ exp\left[ -q^2 \left( x - x_C + N x_1 \right)^2 \right] + exp\left[ -q^2 \left( x + x_C + N x_1 \right)^2 \right] \right\} \qquad (37b)$$

allowing a wide variety of beam profiles, and an arbitrary number of "slits", according to the choice of the parameters $a, b, q, M, x_C , x_1$ .

The values of $R(x; z > 0)$ are then computed step by step by means of a symplectic integration method, and connected, at each step, by a Lagrange interpolation, allowing to perform space derivatives and providing both $G(x; z > 0)$ and the full set of trajectories. We show in Fig.2 and Fig.3, respectively, the initial (continuous) and final (dashed) transverse profiles of the **beam intensity** ($\div R^2(x,z)$, in arbitrary units) and of the **potential function** $G(x,z)$, for the diffraction-like case obtained from eq. (37a) with $a=0, b=1, q=1.68, M=2, x_C=0.31$. Fig. 4 shows the corresponding set of **trajectories** on the (z,x)-plane. Fig.5 and Fig.6 present, respectively, the initial (continuous) and final (dashed) transverse profiles of **beam intensity** and **potential function** $G(x,z)$ for the case obtained from eq. (37b) with $q=3.5$; $M=3$; $x_C=1.15$; $x_1=0.3$, and Fig.7 shows the corresponding set of beam **trajectories** on the (z,x)-plane. Fig.(8) shows the initial transverse profiles of beam **intensity** (continuous line, left scale) and **potential function** G (dashed line, right scale) for the interference of two simple Gaussians obtained from eq.(37a) with $a=0$, $b=1$, $q=1$, $M=1$, $x_C=2.5$. Figs. 9 and Fig.10 present, respectively, the initial (continuous) and final (dashed) transverse profiles of beam **intensity** and **potential function**, choosing the scales in such a way as to evidence the details of the fringe formation, and Fig.11 shows the corresponding set of **beam trajectories** on the (z,x)-plane.

## 5 Discussion

5.1 - For a *stationary potential* V(x,y,z) the *time-independent* Schrödinger equation (20) admits in general, as is well known, a set of *stationary* (discrete or continuous, according to the boundary conditions) eigen-modes and energy eigen-values, which we shall call respectively $u_n(x,y,z)$ and $E_n$, referring for simplicity to the discrete case. If we recall now eqs.(2) and (22) and define the eigen-frequencies $\omega_n \equiv E_n/\hbar$, together with the eigen-waves

$$\psi_n(x,y,z,t) = u_n(x,y,z)e^{-i\omega_n t} \equiv u_n(x,y,z)e^{-i\frac{E_n}{\hbar}t} \qquad (38)$$

and with an arbitrary linear superposition of them,

$$\psi(x,y,z,t) = \sum_n c_n \psi_n , \qquad (39)$$

with constant coefficients $c_n$, such a function, when inserted into the *time dependent* Schrödinger equation (23), reduces it to the form

$$\sum_n c_n e^{-i\frac{E_n}{\hbar}t} \left\{ \nabla^2 u_n + \frac{2m}{\hbar^2}[E_n - V(x,y,z)]\, u_n \right\} = 0 \qquad (40)$$

showing that it provides a general solution of eq.(23) itself. A solution, indeed, whose Born interpretation **[39]** has become one of the basic principles of quantum mechanics, although "no generally accepted derivation of the Born rule has been given to date" **[40]** .

Would such a superposition allow - *within our description* - a further contribution to the analysis of diffraction and/or interference?

In full agreement with the *classical* description of a point-like particle moving in a stationary potential, our trajectory-based *quantum* approach describes each particle of an assigned mono-energetic beam as a classical-looking point-like object starting from an assigned point-like position and following a well defined stationary trajectory, determined *ab initio* by the launching conditions and ruled by a stationary Quantum Potential, up to a well defined, monochromatic diffraction/interference pattern. Representing particles as wave-packets would only replace each trajectory - *within our approach* - by a bundle of diverging trajectories, due to the spread of different frequencies.

We remind once more that *wave mono-chromaticity* warrants the very possibility of defining the concept of "trajectories", i.e. of lines orthogonal to monochromatic wave-fronts.

5.2 - Coming now to the case of a *time dependent potential V(x,y,z,t)*, the standard Ansatz is to *assume*, with questionable plausibility arguments **[41]**, a Schrödinger equation of the form

$$\nabla^2 \psi - \frac{2m}{\hbar^2} V(x,y,z,t) \psi = -\frac{2m\,i}{\hbar} \frac{\partial \psi}{\partial t} \qquad (41)$$

where no solution of the form (39) is generally possible, except for the zero-order approximation of small perturbation cases and for the transition between stationary eigen-states. Eq.(41) is assumed however as a basic working tool, whose final confirmation relies on experiment, even for strong perturbations, in highly specialistic fields such as molecular **[42]** and ultra-fast laser **[43,44]** dynamics.

Would such an equation provide a further contribution to the analysis of diffraction and/or interference?

It is difficult to admit the benefits of a time-varying potential - or, as that, of moving slits - on the observation of diffraction/interference patterns.

We believe therefore that the *stationary media* and *trajectories* and the *monochromatic waves* considered here are the simplest and most reasonable assumption for the family of problems treated in the present work.

## 6 Conclusions

We summarize, in conclusion, the goals reached in the present paper by stating that:

1) a trajectory-based Hamiltonian approach, going much beyond the geometrical optics approximation, was found for the description of typically wave-like stationary features (such as diffraction and/or interference) of *classical* wave beams;

2) a stationary *Wave Potential* function, due to the structure itself of *Helmholtz-like* equations, coupling all the beam trajectories and acting perpendicularly to them, was defined and shown to be a *general wave property*, allowing a deterministic description of *quantum matter waves* as well as *classical waves*, and containing the so-called *Quantum Potential* as a particular case;

3) the numerical integration of the Hamiltonian set of equations (33) was performed, for wave diffraction and interference, in a form holding both for classical and quantum waves and avoiding a wave-packet representation of the moving particles.

**Figures and captions**

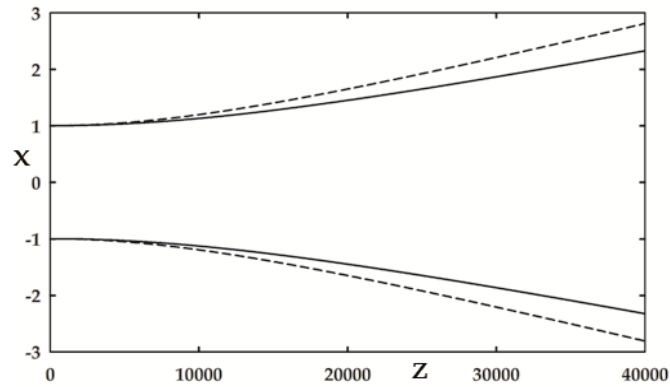

**Fig.1**   Waist trajectories of the Gaussian beams obtained for two different values of $\varepsilon$ : $\varepsilon_1 = 1.65 \times 10^{-4}$  (continuous lines) and   $\varepsilon_2 = 1.25 \times 1.65 \times 10^{-4}$  (dashed lines).

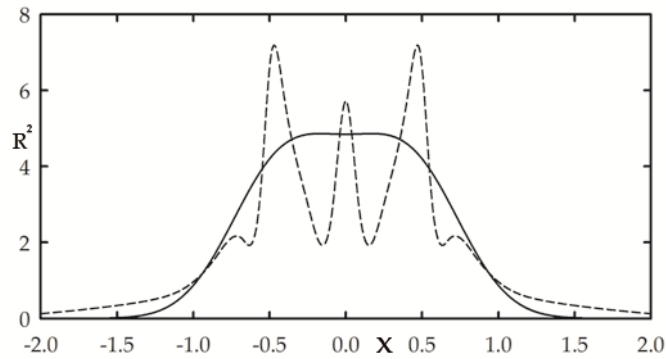

**Fig.2**   Initial (continuous) and final (dashed) transverse profiles of the beam intensity for the diffraction-like case obtained from eq. (37a) with $a = 0, b = 1, q = 1.68, M = 2, x_C = 0.31$ .

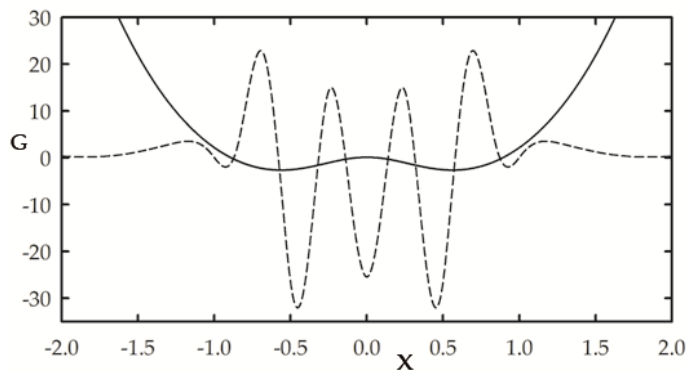

**Fig.3**   Initial (continuous) and final (dashed) transverse profiles of the potential function G corresponding to **Fig.2**.

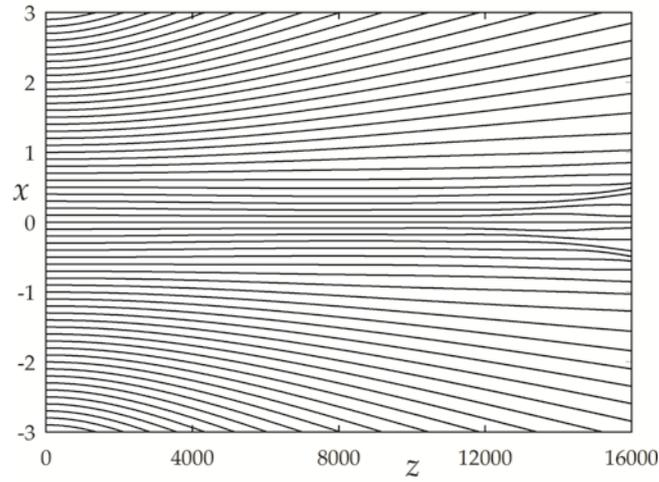

**Fig.4**  Beam trajectories corresponding to Fig.2.

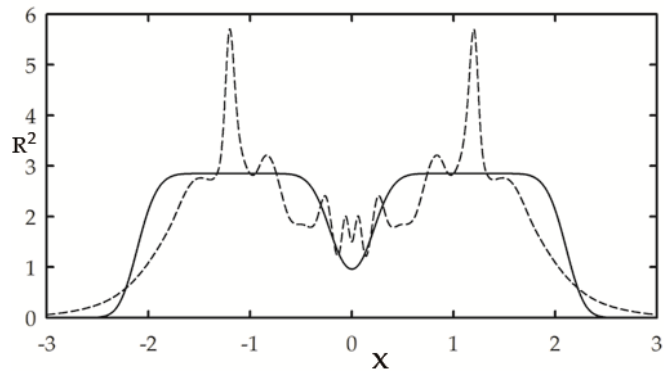

**Fig.5**  Initial (continuous) and final (dashed) transverse profiles of the **beam intensity** for the diffraction-like case obtained from eq. (37b) with $q = 3.5$; $M = 3$; $x_C = 1.15$; $x_1 = 0.3$.

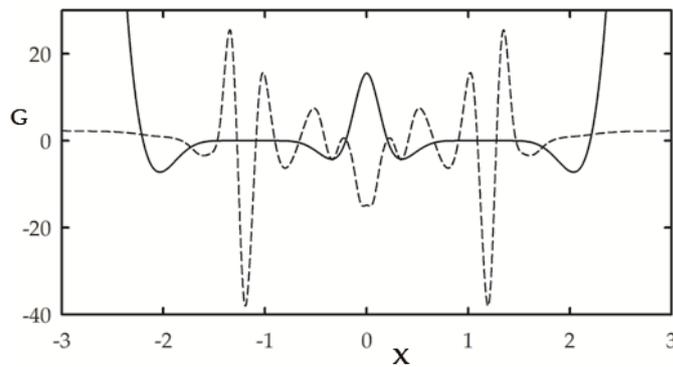

**Fig.6**  Initial (continuous) and final (dashed) transverse profiles of the potential function G corresponding to Fig.5.

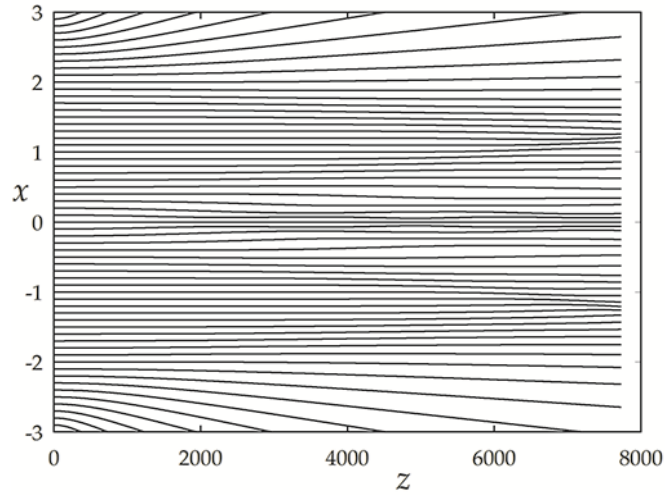

**Fig.7**   Beam trajectories corresponding to Fig.5.

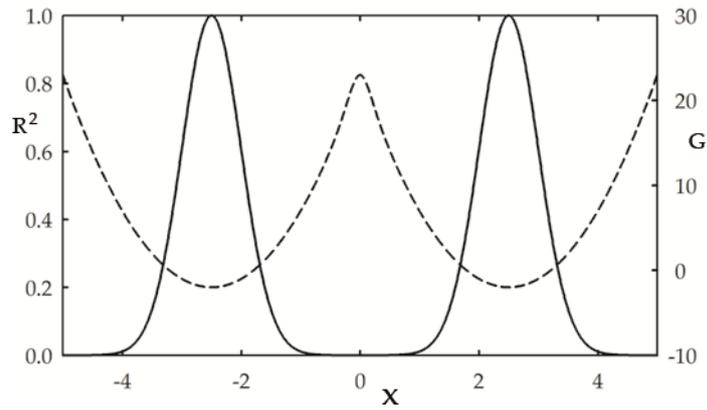

**Fig.8**   Initial transverse profiles of beam intensity (continuous line, left scale) and potential function G (dashed line, right scale) for the interference of two simple Gaussians obtained from eq.(37a) with $a = 0$, $b = 1$, $q = 1$, $M = 1$, $x_C = 2.5$ .

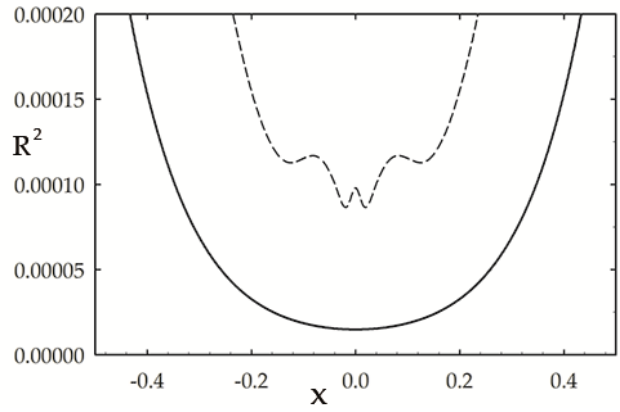

**Fig.9**  Initial (continuous) and final (dashed) transverse profiles of the beam intensity corresponding to Fig.8.

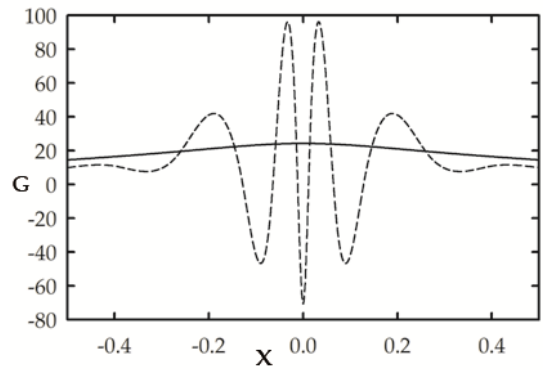

**Fig.10** Initial (continuous) and final (dashed) transverse profiles of the potential function G corresponding to Fig.8.

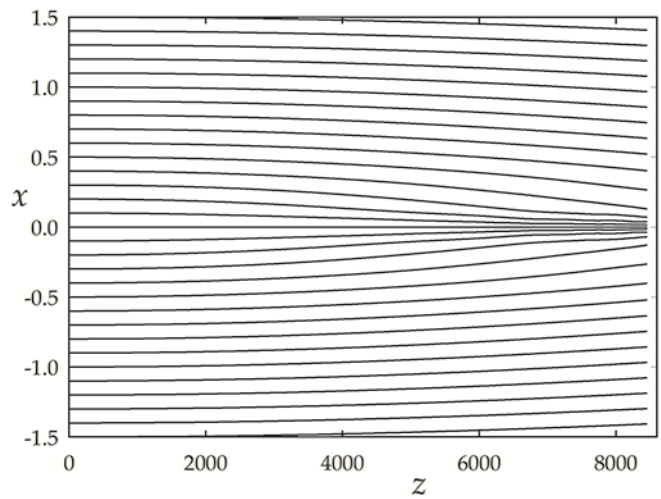

**Fig.11**  Beam trajectories corresponding to Fig.8.